\documentclass[aps,prl,twocolumn,showpacs,10pt,floatfix]{revtex4}
\usepackage{graphicx}
\usepackage{amssymb}
\usepackage{amsmath}

\begin{document}
\title{Theory of Electromagnetically Induced Transparency in Strongly Correlated Quantum Gases}
\author{H. H. Jen$^{1}$}
\author{Daw-Wei Wang$^{1,2}$}
\affiliation{$^{1}$Physics Department, National Tsing Hua University, Hsinchu, Taiwan, R. O. C.\\
$^{2}$Physics Division, National Center for Theoretical Sciences, Hsinchu, Taiwan, R. O. C.}

\newcommand{\p}{\mathbf{p}}
\renewcommand{\r}{\mathbf{r}}
\renewcommand{\k}{\mathbf{k}}
\newcommand{\q}{\mathbf{q}}
\date{\today}
\begin{abstract}
We develop a general theory to study the electromagnetically induced transparency (EIT) in ultracold quantum gases, applicable for both Bose and Fermi gases with arbitrary inter-particle interaction strength. We show that, in the weak probe field limit, the EIT spectrum is solely determined by the single particle Green's function of the ground state atoms, and reflects interesting quantum many-body effects when atoms are virtually coupled to the low-lying Rydberg states. As an example, we apply our theory to 1D Luttinger liquid, Bose-Mott insulator state, and the superfluid state of two-component Fermi gases, and show how the many-body features can be observed non-destructively in the unconventional EIT spectrum.
\end{abstract}
\maketitle
\underline{\it Introduction:} In the last decade, systems of ultracold atoms open up great opportunities to study many-body physics, which may be even not existing in traditional condensed matter systems \cite{manybody}. Huge advances in experimental techniques of tuning interaction strengths and generating optical lattice enable studying strongly correlated atomic gases with high controllability. Most of the experimental measurements rely on the interference pattern of matter wave in different circumstances, for example the time-of-flight experiment, the noise correlations measurement \cite{noise}, and the Bragg scattering spectroscopy \cite{Bragg}, etc. In-situ imaging is also shown important in characterizing the equation of the state and the critical properties near the phase transition point \cite{Chin,Ho}. 

Besides of the many-body problem, the quantum control and manipulation of light-atom interaction is also an extensively studied field in cold atoms. One of the most important examples is the Electromagnetically Induced Transparency (EIT), which has led to several interesting subjects such as the dark-state polariton, slow light, and induced photon-photon interaction in a cold atomic gas \cite{Lukin_rmp,Fleischhauer_rmp}. EIT experiments in quantum degenerate gases \cite{BEC_Hau,Bloch} or in Rydberg states \cite{Rydberg_rev, Ryd_1,Ryd_2,Ryd_3} are also recently explored by several groups. However, so far the EIT theory is still mainly based on the single-particle picture (or within a mean field approximation), and therefore cannot clarify the non-trivial coupling between light-atom interaction and the many-body physics. For example, how is the EIT spectrum modified when the ground state is a strongly correlated state without single particle excitations?\ In what conditions can a non-trivial many-body effect be measured experimentally? Resolving these interdisciplinary problems is not only an interesting theoretical subject itself, but may be also applied to experimentally detect some many-body properties via EIT spectroscopy.
\begin{figure}[t]
\centering\includegraphics[height=4.0cm, width=7.5cm]{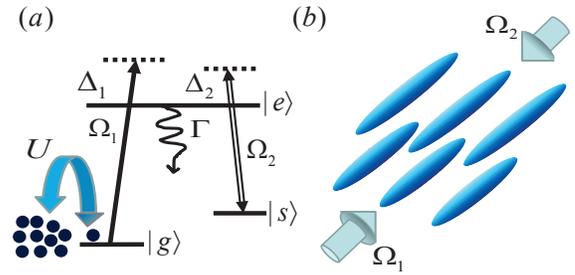}
\caption{(Color online) (a) Schematic figure for a standard EIT experiment in a strongly interacting atomic gas with the $\Lambda$-type scheme:  The control ($\Omega_{2}$) and the probe ($\Omega_1$) fields couple two hyperfine ground states $|g\rangle$ and $|s\rangle$ with the excited state $|e\rangle$ (detunings are $\Delta_2$ and $\Delta_1$ respectively). $\Gamma$ is the spontaneous decay rate of $|e\rangle$. Filled circles denote atoms with mutual interaction $U$ in the ground state. (b) Experimental setup for EIT on a 2D array of 1D Luttinger liquids in the counter-propagating scheme.}%
\label{eitfig}
\end{figure}

In this paper, we develop a general EIT theory in a strongly interacting quantum gas, taking into account the full kinetics and inter-atom interaction, within the weak probe field limit. We explicitly show how the experimental EIT spectrum is directly related to the dynamical Green's function, and the quantum many-body phenomena is manifested most when atoms are coupled to a low-lying Rydberg state due to the sufficient recoil energy compared to the decay rate of the excited state \cite{A}. As an example, we apply our theory to three important strongly correlated systems: Luttinger liquid of 1D Bose gases, Bose-Mott insulator in an optical lattice, and the superfluid state of two-component Fermi gases. For Luttinger liquid, we demonstrate a power law dependence of the EIT spectrum near the resonance, while the significant frequency shift and the asymmetric absorption spectrum can be identified for the other cases. Our results for the first time demonstrate the strongly correlated effects in the EIT experiment, suggesting a non-destructive measurement on the quantum gases.

\underline{\it General EIT theory for quantum gases:} 
We consider the conventional EIT setup ($\Lambda$ type scheme) as shown in Fig. \ref{eitfig}(a): atoms are initially prepared in the ground state $|g\rangle$. The probe ($\Omega_1$) and control ($\Omega_2$) fields couple the ground state to another low energy state $|s\rangle$ and an excited state $|e\rangle$ with detunings $\Delta_1=\omega_1-(E_e-E_g)$ and $\Delta_2=\omega_2-(E_e-E_s)$ respectively where we set $\hbar=1$. Here $\omega_{1,2}$ are central frequencies of probe and control fields, while $E_{g,s,e}$ denote the energies of atoms in different internal states. After using the dipole approximation and rotating wave approximation, the total Hamiltonian ($\hat{H}_{\rm tot}$$=$$\hat{H}_0+\hat{H}_U+\hat{H}_1(t)$) in the rotating frame becomes:
\begin{align}
&\hat{H}_0=\sum_{c=\{g,s,e\}}\sum_\k(\frac{\k^2}{2m}-\mu)\hat{c}_\k^\dagger\hat{c}_\k+\sum_\k[(\Delta_2-\Delta_1)\hat{s}_\k^\dagger\hat{s}_\k\nonumber\\
&-\Delta_1\hat{e}_\k^\dagger\hat{e}_\k]-\sum_\k(\Omega_2\hat{e}_{\k+\k_1}^\dagger\hat{s}_{\k+\k_1-\k_2}+h.c.),\nonumber\\
&\hat{H}_U=\frac{1}{2V}\sum_{c,d=\{g,s,e\}}\sum_{\k,\k',\q}U_{cd}\,\hat{c}_{\k+\q}^\dagger\hat{d}_{\k'-\q}^\dagger\hat{d}_{\k'}\hat{c}_\k,\nonumber\\
&\hat{H}_1(t)=-\frac{1}{V}\sum_{\k,\q}\bar{\Omega}_{1,\k}(t)\hat{e}_{\k+\q}^\dagger\hat{g}_\q+h.c.
\end{align}
Here $\hat{c}_\k$ and/or $\hat{d}_\k$ denote the atomic field operators in the momentum space. $\hat{H}_0$ includes all the single particle part without the probe field for atomic mass $m$, chemical potential $\mu$, and the Rabi frequency of plane wave control field $\Omega_2$. All the inter-atom interactions are included in $\hat{H}_U$ with the short-range interaction $U_{cd}$ (extension to a certain long-ranged interaction is straightforward). Finally, $\hat{H}_1$ denotes the effect of the probe field with $\bar{\Omega}_{1,\k}(t)$ being the slow-varying Rabi frequency of wavevector $\k$. $V$ is the quantization volume and $\k_{1,2}$ are momenta of the probe and control fields respectively.

To derive the electric susceptibility in the linear response of probe field \cite{Fleischhauer_rmp}, we treat $H_1$ as a time-dependent perturbation and keep the full interaction effects.  Therefore, $\hat{H}\equiv \hat{H}_0+\hat{H}_U$ can be separated into two decoupled parts: $\hat{H}=\hat{H}^{g}+\hat{H}^{se}$, where $\hat{H}^{g}$ includes all the ground state kinetic energy and interaction energy terms, while $\hat{H}^{se}$ includes all the single particle terms of states $|s\rangle$ and $|e\rangle$ (see Supplemental material (SM) \cite{SM}).  In the leading order limit of a weak probe, all atoms are in the ground state ($|g\rangle$) and none in states $|e\rangle$ and $|s\rangle$. As a result, the interaction between $|g\rangle$ and the other two states are nothing but a background energy shift in the single particle energy, i.e. $E_{s/e}\to E_{s/e}+nU_{g,s/e}$, where $n$ is the total particle density. The mutual interaction between $|s\rangle$ and $|e\rangle$ are the second order effect and hence negligible. Similar treatment can be also easily applied to long-ranged interaction if Rydberg states of high principal quantum number are involved \cite{B}.

A phenomenonological spontaneous decay rate ($\Gamma$) of the excited state can be added, and we also assume negligible dephasing rate between the two hyperfine ground states. Note that when considering standard D2 transition, $\Gamma\approx 6$ MHz, is much larger than the average atomic kinetic energy, and therefore the quantum many-body effect may not be easily observed.

Defining the polarization operator, $\hat{P}(\r,t)$ $=$ $d_0[\hat{\psi}_e^\dagger(\r,t)\hat\psi_g(\r,t)+h.c.]$, where $\hat{\psi}_{g/e}^\dagger(\r,t)$ is the field operator, and $d_0$ is the dipole moment, we calculate its variation with respect to the probe field ($H_1$) via the linear response theory \cite{manyparticle}: $\delta\langle\hat{P}(\r,t)\rangle$ $=$ $i\int_{-\infty}^{t}dt'$ ${}_H\langle\Psi_G|[\hat{H}_{1,H}(t'),\hat{P}_H(\r,t)]|\Psi_G\rangle_H$, where $\hat{H}_{1,H}(t)$ and $\hat{P}_H(\r,t)$ are corresponding operators in Heisenberg picture, and $|\Psi_G\rangle_H$ is the ground state of the unperturbed Hamiltonian $\hat{H}=\hat{H}^{g}+\hat{H}^{se}$. In the momentum-frequency space, the electric susceptibility can be obtained to be (see SM \cite{SM}):
\begin{align}
&\chi(\q,\omega)=-\frac{d_0}{V}\sum_\k\int_{-\infty}^{\infty}d\tilde{\omega}i\tilde{G}^<(\k,\tilde{\omega})\nonumber\\
&\Big[\frac{\cos^2\phi_{\k+\q}}{\tilde{\omega}-\omega-\epsilon_{-}(\k+\q)}+\frac{\sin^2\phi_{\k+\q}}{\tilde{\omega}-\omega-\epsilon_{+}(\k+\q)}\Big].\label{chi}
\end{align}
Here $\tilde{G}^{<}(\k,\tilde{\omega})$ is the Fourier transform of the Green's function at zero temperature \cite{manyparticle}:  $G^<(0,0;\r,t)\equiv\mp i{}_H\langle\Psi_G|\hat{\psi}_g^\dagger(\r,t)\hat{\psi}_g(0,0)|\Psi_G\rangle_H\theta(t)$, where the upper (lower) sign refers to bosons (fermions). We note that the above result is very general in the limit of weak probe field, and all the many-body effects are hidden in $\tilde{G}^{<}(\k,\tilde{\omega})$. However, as pointed out in Ref. \cite{Jen}, such many-body effect can be identified more easily in the counter-propagating (rather than the co-propagating) setup of the EIT experiment, since the recoil momentum of the former is manifested more significantly in the quantum statistics. In the rest of this paper, therefore, we just consider the counter-propagating scheme in order to measure the many-body effect from the EIT spectrum.

\underline{\it EIT in Luttinger liquid:} Now we study EIT in a Luttinger liquid (LL) as shown in Fig. \ref{eitfig}(b), which is a very general 1D effective model \cite{Haldane,LL} and has no condensate even at zero temperature. Since all the elementary excitations of a LL are collective, it is therefore interesting to investigate how the probe field propagates inside such strongly correlated system. The single particle Green's function can be exactly calculated by bosonization method \cite{LL} (see SM \cite{SM}): 
\begin{align}
iG^<_{LL}(x,t)=\frac{na^{1/2\kappa}}{\big[x^2+(a+ivt)^2\big]^{1/4\kappa}},
\label{G_LL}
\end{align}
where $\kappa$ is the Luttinger parameter, $v$ is the phonon velocity, and $a^{-1}$ is the system-dependent momentum cut-off. $1<\kappa<\infty$ for a short-ranged repulsive interaction, while $\kappa$ can be smaller than one if the interaction is long-ranged.

\begin{figure}[t]
\centering\includegraphics[height=4.5cm, width=8.5cm]{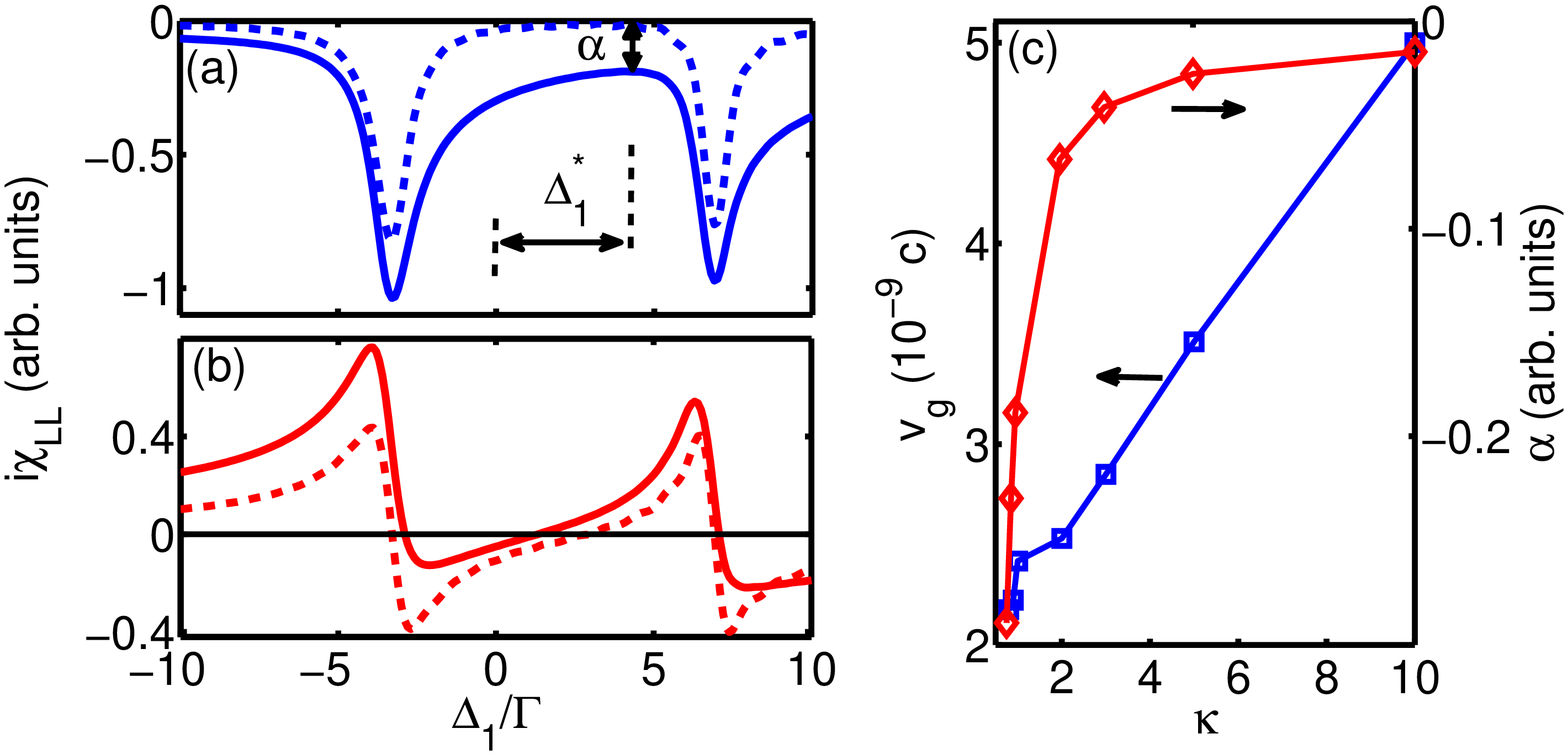}
\caption{(Color online) The EIT profiles for a Luttinger liquid of $^{87}$Rb atoms in the counter-propagating excitation scheme. We take a static probe field ($\q,\omega=0$) and a resonant control field ($\Delta_2=0$) with Rabi frequency $\Omega_2=5~\Gamma$. The excited state is chosen as low-lying Rydberg transition of $|24\textrm{P}_{3/2}\rangle$ with the spontaneous decay rate, $\Gamma^{-1}=28.3~\mu\text{s}$. The effective 1D interaction strength can be tuned by the confinement resonance of 1D tubes \cite{1D_confinement}. (a) Absorption (Re[$i\chi$]) and (b) dispersion (Im[$i\chi$]) profiles for $\kappa=1$ (solid) and $10$ (dash). The horizontal line guides the eye to the zero in (b).\ (c) shows the group velocity $v_g$ ($\square$), and the absorption depth $\alpha$ ($\Diamond$) at the transparency point ($\Delta_1^*$) as a function of $\kappa$. For convenience, we take the phonon velocity $v$=4.3 mm/s and the cut-off $a$=0.12 $\mu$m within a typical experimental regime \cite{1D}. $c$ is the speed of light.}
\label{LL}
\end{figure}

In Figs. \ref{LL}(a) and (b), we show the full numerically integrated EIT absorption and dispersion profiles (from Eq. (\ref{chi}) and (\ref{G_LL})) of a 1D $^{87}$Rb atoms (see the caption for parameters) in the counter-propagating scheme. It is easy to see that the power law dependence of $\chi_{LL}$ makes the dispersion highly asymmetric about the resonance points and the absorption depth ($\alpha$) becomes larger at the transparency point ($\Delta_1^*$) in the strong interacting regime (smaller $\kappa$), in contrast to the standard EIT profiles in the weak interacting limit ($\kappa\gg 1$). This is because when a ground state atom is excited by the probe field, the transition matrix elements are composite of all the collective excitations of different energies, weighted by the density of states in a power-law distribution. In Fig. \ref{LL}(c), we show the group velocity of light (proportional to the slope of dispersion at the transparent point, where the absorption is the smallest) as a function of $\kappa$. One can see that for a stronger interaction (smaller $\kappa$), the group velocity becomes smaller due to steeper dispersion relations, but the absorption is also larger, making the slow light propagate less effectively in a LL. On the other hand, such unconventional EIT spectrum also implies a sensitive measurement about the many-body properties.

\underline{\it EIT in a Mott insulator:}
Now we consider the Mott-Insulator (MI) of strongly interacting bosons in a 3D optical lattice, which can be well described by a single band Hubbard model (HM) \cite{BHM}. The ground state field operator is expressed as $\hat{\psi}_g(\r,t)=\sum_\mathbf{R} \hat{g}_\mathbf{R}(t) w_\mathbf{R}(\r)$ where $w_\mathbf{R}(\r)$ is the Wannier function, and $\hat{g}_\mathbf{R}(t)$ is the field operator at site ${\bf R}$. When deep inside the MI state, we can use the three-state model \cite{three,Mott} to control the small number fluctuation above the Fock state, and obtain the Green's function at zero temperature (see SM \cite{SM}): $i\tilde{G}^<(\k,t)=n_0\sum_{\mathbf{R}}|\tilde{w}_{\mathbf{R}}(\k)|^2e^{-i\epsilon_h(\k)t}\theta(t)$, 
where $n_0$ is the integer filling fraction, $\tilde{w}_{\mathbf{R}}(\k)$ is the Fourier transform of $w_\mathbf{R}(\r)$, and $\epsilon_h(\k)$ $=$ $\epsilon_0(\k)/2+\delta\mu+\tilde{\omega}(\k)$ is the hole excitation with $\tilde{\omega}(\k)\approx U/2$ when deep inside the Mott state. Here $\epsilon_0(\k)\equiv 2J\sum_{\alpha=1}^{3}\cos(\k_\alpha d)$, and $\delta\mu=-3J$ for a 3D square lattice of lattice constant $d$. $J$ and $U$ are the tunnelling amplitude and on-site interaction.\ As a result, the susceptibility from Eq. (\ref{chi}) becomes
\begin{align}
\chi_{MI}(\q,\omega)&
=\frac{d_0N}{V}\sum_\k|\tilde{w}_\mathbf{R}(\k)|^2\bigg[\frac{\cos^2\phi_{\k+\q}}{\omega+\epsilon_{h}(\k)+\epsilon_{-}(\k+\q)}\nonumber\\
&+\frac{\sin^2\phi_{\k+\q}}{\omega+\epsilon_{h}(\k)+\epsilon_{+}(\k+\q)}\bigg].
\label{Motteq}
\end{align}

\begin{figure}[t]
\centering\includegraphics[height=4cm, width=9cm]{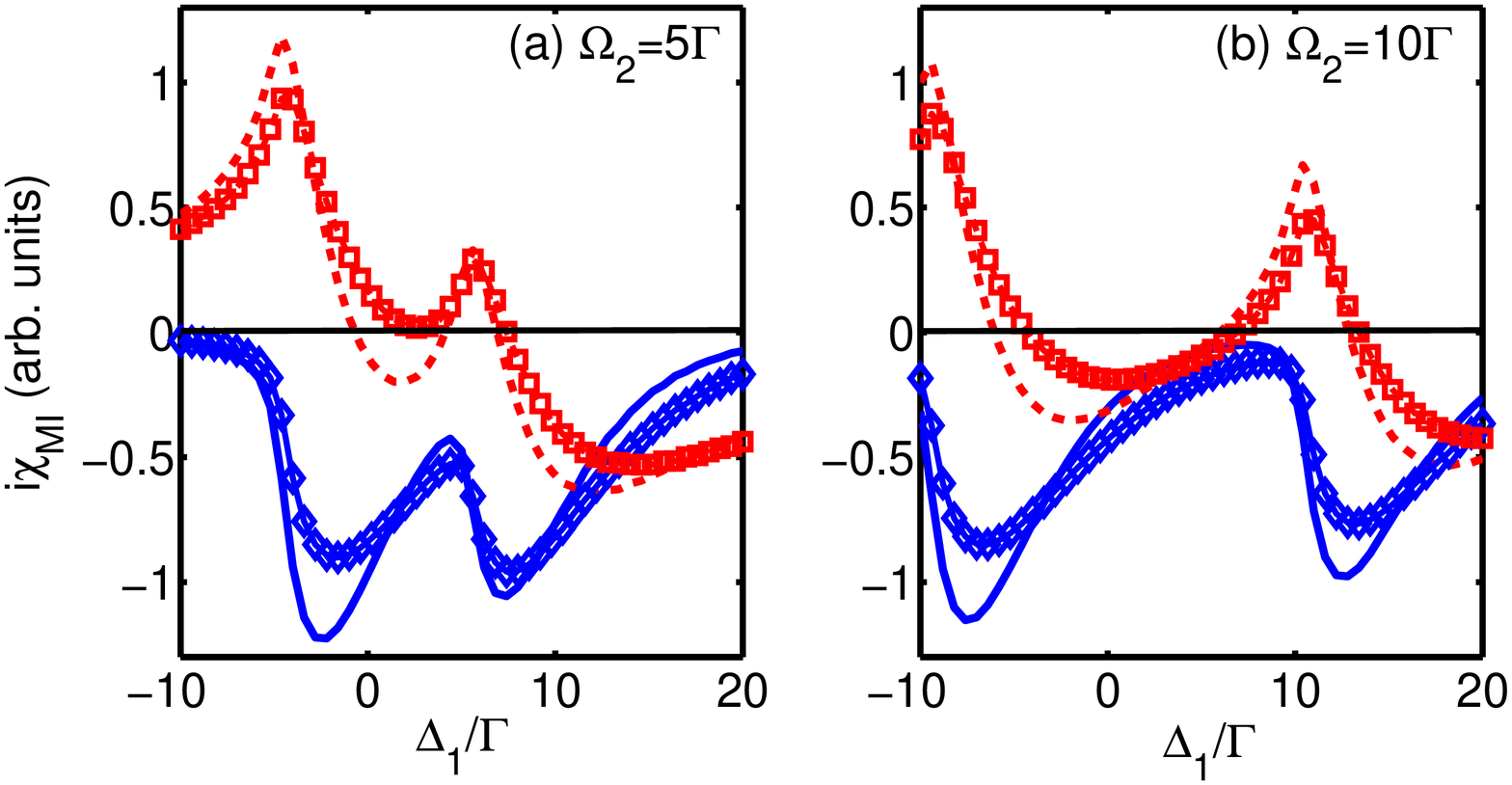}
\caption{(Color online) EIT profiles of a unit-filling Mott-insulator with $^{87}$Rb atoms loaded in a 3D optical lattice (lattice constant $d=426$ nm). The corresponding parameters ($U$, $J$) in unit of the recoil energy $E_R$ are (0.5, 0.007) and (1.05, $10^{-4}$) for $V_0/E_R=15$ (solid-blue and dash-red) and $40$ ($\Diamond$ and $\square$) respectively.\ Absorption (Re[$i\chi$],solid-blue, $\Diamond$) and dispersion (Im[$i\chi$],dash-red, $\square$) profiles are plotted for various control field strengths (a) $\Omega_2=5$, and (b) $10~\Gamma$. All other parameters are the same as Fig. 2.}%
\label{Mottfig}
\end{figure}

Compared to the standard EIT spectrum, we find that $\chi_{MI}$ is contributed by EIT in different momentum states, weighting by the Wannier function of the optical lattice. When the momentum distribution is large in a strong lattice strength, the contribution from different momenta makes the EIT spectrum much broadened, in contrast to the superfluid state which is similar to the standard EIT spectrum in a Bose-Einstein condensate (BEC) \cite{Jen}.

In Fig. \ref{Mottfig}, we demonstrate the EIT spectrum inside the MI state of $^{87}$Rb with unit filling for different control fields and lattice strengths. As discussed above, the momentum distribution of Wannier function lead to an asymmetric and inhomogeneous broadening of the absorption profiles near the resonance. It makes the EIT window less transparent when the control field is weak (Fig. \ref{Mottfig}(a)), while it becomes less effective in the large field limit. In other words, by measuring the absorption depth one may also determine the momentum distribution of the underlying many-body system.

\underline{\it EIT in a BCS superfluid state:}
Finally we investigate the EIT in a BCS superfluid of two-component Fermi gases, using similar setup \cite{pairing} by coupling one of the atomic ground state to a low-lying Rydberg state \cite{A}.\ Without losing generality, here we just presume the existence of a superconducting gap, $\Delta_{BCS}$, not specifying the pairing mechanism. One of the two components (pseudo-spin up) is our ground state ($|g\rangle$) in the EIT experiment, and the other is not involved in the EIT process. The Green's function can be calculated by transforming toward the Bogoliubov quasi-particles ($\alpha$ and $\beta$): $\hat{g}_{\k,\uparrow}=\cos\theta_\k\hat{\alpha}_\k+\sin\theta_\k\hat{\beta}_{-\k}^\dagger$, where $\sin^2\theta_\k\equiv(1-\xi_\k/E_\k)/2$. Here $E_\k=\sqrt{\Delta_{BCS}^2+\xi_\k^2}$ is the excitation energy with $\xi_k\equiv\k^2/(2m)-\mu$ and $\mu$ being the chemical potential. We find that
$-i\tilde{G}^<_{BCS}(\k,t)=\sin^2\theta_\k^2e^{-iE_\k t}\theta(t)$ (See SM \cite{SM}), and then
\begin{align}
\chi_{BCS}(\q,\omega)&=\frac{d_0}{V}\sum_\k \sin^2\theta_\k \Big(\frac{\cos^2\phi_{\k+\q}}
{\omega+E_\k+\epsilon_{-}(\k+\q)}\nonumber\\&
+\frac{\sin^2\phi_{\k+\q}}{\omega+E_\k+\epsilon_{+}(\k+\q)}\Big).
\end{align}

Note that in the limit of zero pairing gap, $\Delta_{BCS}=0$, $\sin^2\theta_\k$ becomes a step function at Fermi energy, and hence $\chi_{BCS}$ becomes the result of a noninteracting Fermi gas \cite{Jen}. 

\begin{figure}[t]
\centering\includegraphics[height=4.7cm, width=8.5cm]{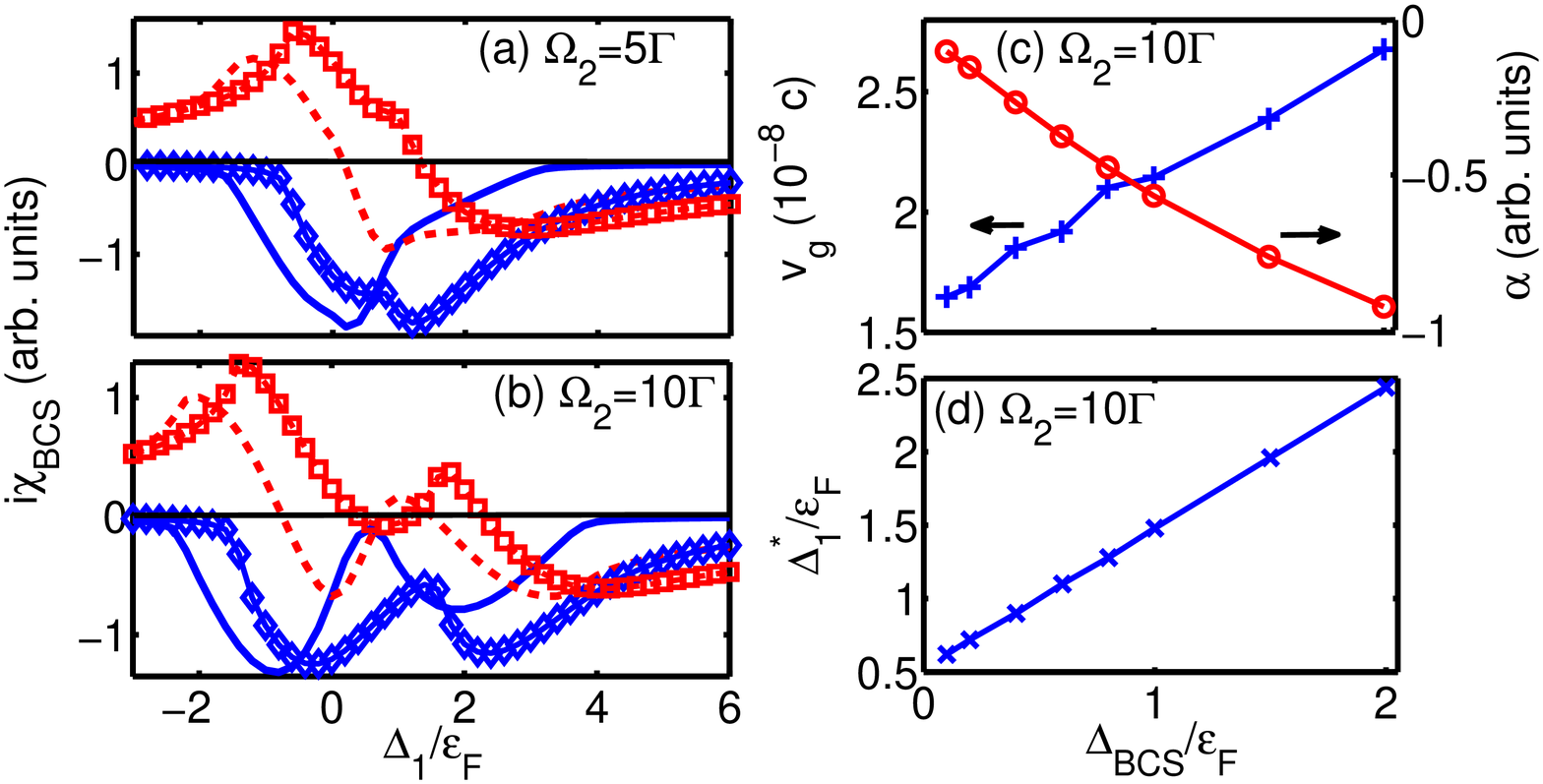}
\caption{(Color online) EIT profiles of a BCS superfluid state of two-component Fermi gases ($^{40}$K).\ The excited state is chosen as low-lying Rydberg transition of $|21\textrm{P}_{3/2}\rangle$ with $\Gamma^{-1}=25.4~\mu\text{s}$. We fix the chemical potential $\mu=\epsilon_F=6.7$ $\hbar\Gamma$ with atomic density $n\sim 10^{14}$ cm$^{-3}$ and resonant control field ($\Delta_2=0$). Absorption (Re[$i\chi$], solid-blue, $\Diamond$) and dispersion (Im[$i\chi$], dash-red, $\square$) profiles are plotted for pairing gap $\Delta_{\rm BCS}/\epsilon_F=0.1$ (solid-blue and dash-red) and $1$ ($\Diamond$ and $\square$), while the control field strengths are (a) $\Omega_2=5$ and (b) $10$ $\Gamma$. (c) shows the group velocity $v_g$ ($+$) and the absorption depth $\alpha$ ($\bigcirc$) at the transparency point as a function of  $\Delta_{\rm BCS}$. (d) shows the transparency positions $\Delta_1^*$ ($\times$).}%
\label{SCfig}
\end{figure}

In Figs. \ref{SCfig}(a) and (b), we plot the EIT profiles of the BCS superfluid phase for $^{40}$K in the counter-propagating excitation schemes. Different control field strengths and different gap amplitudes are shown together for comparison. One can see that when the control field is turned smaller, the recoil energy becomes more important so that the presence of Fermi sea makes the EIT spectrum disappeared, similar to the noninteracting fermions \cite{Jen}. When the control field becomes stronger, the quantum degeneracy becomes less important so that the standard EIT features appear, while an asymmetric broadening appears when the gap is opened. Interestingly, we find the group velocity is larger for a larger $\Delta_{BCS}$ (see Fig. 4(c)),  different from results of a LL. In (d) we show that the transparency points $\Delta_1^*$ scales as the gap energy, and therefore can be used to measure the pairing gap.

\underline{\it Experimental issues:}
The difference between EIT approach and standard Bragg spectroscopy (BS) has two folds.  Firstly EIT is nondestructive while BS operates with interference of atoms.  Secondly BS measures the density-density correlation function, in contrast to the first order correlation in EIT. For experimental realization, it may involve the integration time and precision of the spectrum measurement.  With kHz resolution required in our scheme, we need ms integration time compatible with the atomic lifetime of several ms, which can be fulfilled by minimizing the inelastic collisions of BEC \cite{coherence_Hau}.\ The contrast of susceptibility can be also enhanced by increasing the optical depth of atoms. Including the thermal excitations is crucial for practical implementations and will be discussed in the future work.

In summary, we develop a general and analytic theory to investigate the EIT spectrum of quantum degenerate ultracold atoms. The spectrum is solely determined by the single particle Green's function, and has most significant quantum many-body physics when the atoms are coupled to a low-lying Rydberg state. In the regime of strong interaction, the absorption profile has a highly asymmetric inhomogeneous broadening with a frequency shift in detuning. Our results suggest a new non-destructive method to investigate the strongly correlated physics from the well-developed EIT experiment.

We appreciate fruitful discussions with Ite A. Yu, Y.-C. Chen, M. Fleischhauer, and G. Juzeli\={u}nas. This work is supported by NSC and NCTS grants in Taiwan.


\clearpage
\appendix
\section{Supplemental Material:}

\renewcommand{\r}{\mathbf{r}}
\renewcommand{\k}{\mathbf{k}}
\renewcommand{\theequation}{S\arabic{equation}}

\section*{Linear response theory for the EIT spectrum}


We apply the linear response theory \cite{manyparticle} to the electric susceptibility of EIT.\ For the perturbation Hamiltonian $\hat{H}_1(t)$, the linear response of any operator $\hat{O}$ under the ground state $|\Psi_G\rangle_H$ of the unperturbed Hamiltonian $\hat{H}$ is ($\hbar\equiv 1$)
\begin{align}
\delta\left\langle \hat{O}(t)\right\rangle  & =i\int_{-\infty}^{t}dt^{\prime}{\Big.}_H\left\langle \Psi_{G}\left\vert \left[  \hat{H}_{1,H}(t^{\prime
}),\hat{O}_{H}(t)\right]  \right\vert \Psi_{G}\right\rangle_{H},\label{linear}
\end{align}
where $\hat{H}_{1,H}(t)=e^{i\hat{H}t}\hat{H}_{1}(t)e^{-i\hat{H}t}$ is the operator defined in the Heisenberg picture (with subscript $H$), and the same as for $\hat{O}_{H}(t)$.

To calculate the electric susceptibility of EIT, we consider the variation of the polarization operator,
\begin{align}
&\hat{P}(\k,t)\equiv\int d\r\hat{P}(\r,t)e^{-i\k\cdot\r}\nonumber\\&=d_{0}\sum_{\q}\left[  \hat{e}_{-\k+\q}^{\dag}(t)\hat{g}_{\q}(t)+\hat{g}
_{\q}^{\dag}(t)\hat{e}_{\k+\q}(t)\right]  ,
\end{align}
and put the above into Eq. (\ref{linear}), we have
\begin{widetext}
\begin{align}
&  \delta\left\langle \hat{P}(\q,t)\right\rangle =i\int_{-\infty}^{\infty}dt^{\prime}\theta(t-t^{\prime}){\Bigg.}_H\left\langle \Psi_{G}\left\vert \left[-\frac{1}{V}\sum_{\k,\k'}e^{i\hat{H}t^{\prime}}\left(\bar{\Omega}_{1,\k}(t^{\prime})\hat{e}_{\k+\k'}^{\dag}(t^{\prime})\hat{g}_{\k'}(t^{\prime})+h.c.\right)e^{-i\hat{H}t^{\prime}},\right.\right.\right.
\nonumber\\&\left.\left.\left.d_{0}\sum_{\q^{\prime}}e^{i\hat{H}t}\left(\hat{e}_{-\q+\q^{\prime}}^{\dag}(t)\hat{g}_{\q^{\prime}}(t)+\hat{g}_{\q^{\prime}}^{\dag}(t)\hat{e}_{\q+\q^{\prime}}(t)\right)  e^{-i\hat{H}t}\right]\right\vert \Psi_{G}\right\rangle_H ,\nonumber\\&  =-i\int_{-\infty}^{\infty}dt^{\prime}\theta(t-t^{\prime})\frac{d_{0}}{V}\sum_{\k,\k',\q^{\prime}}\Big\{\bar{\Omega}_{1,\k}(t^{\prime}){\Big.}_H\left\langle \Psi_{G}\left\vert\left[e^{i\hat{H}t^{\prime}}\hat{e}_{\k+\k'}^{\dag}(t^{\prime})\hat{g}_{\k'}(t^{\prime})e^{-i\hat{H}t^{\prime}},e^{i\hat{H}t}\hat{g}_{\q^{\prime}}^{\dag}(t)\hat{e}_{q+\q^{\prime}}(t)e^{-i\hat{H}t}\right]  \right\vert
\psi_{G}\right\rangle_H \nonumber\\
&  +\bar{\Omega}_{1,\k}^{\ast}(t^{\prime}){\Big.}_H\left\langle \psi_{G}\left\vert\left[  e^{i\hat{H}t^{\prime}}\hat{g}_{\k'}^{\dag}(t^{\prime})\hat{e}
_{\k+\k'}(t^{\prime})e^{-i\hat{H}t^{\prime}},e^{i\hat{H}t}\hat{e}_{-\q+\k^{\prime}}^{\dag}(t)\hat{g}_{\q^{\prime}}(t)e^{-i\hat{H}t}\right]
\right\vert \Psi_{G}\right\rangle_H\Big\},
\end{align}
\end{widetext}
where in the last line we have neglected the terms of $\hat{e}^{\dag}\hat{g}\hat{e}^{\dag}\hat{g}$ and $\hat{g}^{\dag}\hat{e}\hat{g}^{\dag}\hat{e}$, which are vanishing in the expectation value of ground state wavefunction, $|\Psi_G\rangle_H$.

For the unperturbed Hamiltonian (or probe-free Hamiltonian, $\hat{H}=\hat{H}_0+\hat{H}_U$), all atoms are in the atomic ground state $|g\rangle$, and states $|e\rangle$ and $|s\rangle$ are empty. Therefore, further expansion of the above equation gives
\begin{widetext}
\begin{align}
& \delta\left\langle \hat{P}(\q,t)\right\rangle=\frac{id_{0}}{V}\int_{-\infty}^{\infty}dt^{\prime}\theta(t-t^{\prime})
\sum_{\k,\k',\q^{\prime}}\Big[  \tilde{\Omega}_{1,\k}(t^{\prime}){\Big.}_H\left\langle\psi_{G}\left\vert \hat{g}_{\q^{\prime}}^{\dag}(t)\hat{e}_{\q+\q^{\prime}}(t)\hat{e}_{\k+\k'}^{\dag}(t^{\prime})\hat{g}_{\k'}(t^{\prime})\right\vert
\psi_{G}\right\rangle _{H} -h.c.(\q\rightarrow-\q)\Big]  .
\end{align}
\end{widetext}

Since the excited state $|e\rangle$ and the second ground state $|s\rangle$ are coupled by the control field, we then replace $\hat{e}_{\k+\k_1}(t)$ with the new eigen-bases $\hat{a}_\k$ and $\hat{b}_\k$ with corresponding eigenenergy $\epsilon_\pm(\k)$ respectively. Here 
$\hat{a}_{\k}$ $=$ $\cos\phi_\k\hat{e}_{\k+\k_1}$ $+$ $\sin\phi_\k\hat{s}_{\k+\k_1-\k_2}$ and $\hat{b}_{\k}$ $=$ $\sin\phi_\k\hat{e}_{\k+\k_1}$ $-$ $\cos\phi_\k\hat{s}_{\k+\k_1-\k_2}$, where $\cos\phi_\k$$=$$\sqrt{[\epsilon_+(\k)-\epsilon_{0,\k+\k_1}+\Delta_1]/[\epsilon_+(\k)-\epsilon_-(\k)]}$. The associated eigenvalues are: $\epsilon_\pm(\k)$$=$$-\Delta_1+[\bar{\Delta}_2+\epsilon_{0,\k+\k_1}\pm\sqrt{(\bar{\Delta}_2-\epsilon_{0,\k+\k_1})^2+4\Omega_2^2}]/2$, where $\bar{\Delta}_2\equiv\Delta_2+\epsilon_{0,\k+\k_r}$, $\epsilon_{0,\k}\equiv\k^2/(2m)-\mu$, and $\k_r\equiv\k_1-\k_2$ as the recoil momentum. A phenomenonological spontaneous decay rate ($\Gamma$) of the excited state can be added by replacing $\epsilon_{0,\k+\k_1}$ with $\epsilon_{0,\k+\k_1}-i\Gamma$.\ After shifting the momentum, $\q\rightarrow\q-\k_1$ and $\k\rightarrow\k-\k_1$, we have
\begin{widetext}
\begin{align}
& \delta\left\langle \hat{P}(\q,t)\right\rangle =\frac{id_{0}}{V}\int_{-\infty}^{\infty}dt^{\prime}\theta(t-t^{\prime})\sum_{\k,\k',\q^{\prime}}\Big\{
\bar{\Omega}_{1,\k}(t^{\prime}){\big.}_H\left\langle\Psi_{G}\right\vert\hat{g}_{\q^{\prime}}^{\dag}(t)\left[\cos\phi_{\q+\q^{\prime}}\hat{a}_{\q+\q^{\prime}}(t)-\sin\phi_{\q+\q^{\prime}}\hat{b}_{\q+\q^{\prime}}(t)\right] \nonumber\\& 
\left[\cos\phi_{\k+\k'}\hat{a}_{\k+\k'}^{\dag}(t^{\prime})-\sin\phi_{\k+\k'}\hat{b}_{\k+\k'}^{\dag}(t^{\prime})\right]
\hat{g}_{\k'}(t^{\prime})\left\vert\Psi_{G}\right\rangle _{H} -h.c.(\q\rightarrow-\q)\Big\},\nonumber\\
& =\frac{id_{0}}{V}\int_{-\infty}^{\infty}dt^{\prime}\theta(t-t^{\prime})\sum_{\k}\Big\{\bar{\Omega}_{1,\q}(t^{\prime})F_{\k}(t-t^{\prime})
\left[\cos^{2}\phi_{\k+\q}e^{-i\epsilon_-(\k+\q)(t-t^{\prime})}+\sin^{2}\phi_{\k+\q}e^{-i\epsilon_+(\k+\q)(t-t^{\prime})}\right] 
\nonumber\\& -h.c.(\q\rightarrow-\q)\Big\},
\end{align}
\end{widetext}
where we have defined the ground state correlation function as $F_{\k}(t-t^{\prime})\delta_{\k,\k^{\prime}}$$=$${}_H\langle \Psi_{G}\vert
\hat{g}_{\k^{\prime}}^{\dag}(t)\hat{g}_{\k}(t^{\prime})\vert \Psi_{G}\rangle_{H}$.\ Use the Fourier transform $\delta\langle \hat{P}(\q,\omega)\rangle$$=$$\int_{-\infty}^{\infty}dt\delta\langle \hat{P}(\q,t)\rangle e^{-i\omega t}$ , and define the electric susceptibility as $\chi(\q,\omega)=\delta\langle\hat{P}(\q,\omega)\rangle/\tilde{\Omega}_{1,\q}(\omega)$ where $\tilde{\Omega}_{1,\q}(\omega)$ is the Fourier transform of $\bar{\Omega}_{1,\q}(t)$, we have (after the change of variables, $t=t^{\prime}+t^{\prime\prime}$, $dt=dt^{\prime\prime}$)
\begin{align}
&\chi(\q,\omega)\nonumber\\&=\frac{id_{0}}{V}\sum_{\k}\int_{-\infty}^{\infty}dt\theta(t)F_{\k}(t)
\left[  \cos^{2}\phi_{\k+\q}e^{-i(\omega+\epsilon_-(\k+\q))t}\right.\nonumber\\&\left.+\sin^{2}\phi_{\k+\q}e^{-i(\omega+\epsilon_+(\k+\q))t}\right].
\end{align}

Finally, since the correlation function $F_\k(t-t')$ is related to the time-ordered retarded Green's function via $\theta(t)F_{\k}(t)=\int d\r e^{i\k\cdot\r}iG^{<}(0,0;\r,t)$, where $iG^{<}(0,0;\r,t)\equiv{}_H\langle\Psi_G|\hat{\psi}_{g}^{\dag}(\r,t)\hat{\psi}_{g}(0,0)|\Psi_G\rangle_H\theta(t)$, we can rewrite above results to be
\begin{align}
&  \chi(\q,\omega)=\frac{id_{0}}{V}\sum_{\k}\int_{-\infty}^{\infty}d\tilde{\omega}i\tilde{G}^{<}(\k,\tilde{\omega})\nonumber\\
&\left[  \frac{i\cos^{2}\phi_{\k+\q}}{\tilde{\omega}-\omega-\epsilon_-(\k+\q)}
+\frac{i\sin^{2}\phi_{\k+\q}}{\tilde{\omega}-\omega-\epsilon_+(\k+\q)}\right],\label{chi1}
\end{align}
where $\tilde{G}^{<}(\k,\tilde{\omega})$ is the Fourier transforms of the Green's
function $iG^<(0,0;\r,t)$.\ The eigenvalues $\epsilon_\pm(\k)$ and $\phi_{\k}$ have been defined before.

As an example we consider a weakly interacting condensate.\ By separating the fluctuation from the condensate: $\hat{g}_\k(t)$ $=$  $\sqrt{N_c}\delta_{\k=0}$ $+$ $\delta\hat{g}_\k(t)$ with $N_c$ being the condensate particle number, we apply Bogoliubov transformation to calculate the single particle Green's function.
The final EIT spectrum can be separated into the condensate (C) and the non-condensate (NC) parts: $\chi(\q,\omega)=\chi_C(\omega)+\chi_{NC}(\q,\omega)$, where 
$\chi_C(\omega)=d_0 n_c[\cos^2\phi_0/(\omega+\epsilon_{-}(0))+\sin^2\phi_0/(\omega+\epsilon_{+}(0))]$, and $\chi_{NC}(\omega)=d_0V^{-1}\sum_{\k\neq 0}v_\k^2[\cos^2\phi_\k/(\omega+\epsilon_{-}(\k)+\epsilon(\k))+\sin^2\phi_\k/(\omega+\epsilon_{+}(\k)+\epsilon(\k))]$.
Here $n_c\equiv N_c/V$ is condensate density, $\epsilon(\k)\equiv(\epsilon_1^2(\k)-\epsilon_2^2)^{1/2}$ is the phonon excitation energy, and $v_\k^2\equiv\sinh^2\theta(\k)$, where $\tanh2\theta(\k)\equiv\epsilon_2/\epsilon_1(\k)$, $\epsilon_1(\k)\equiv\k^2/2m+n_cU_{gg}$, and $\epsilon_2\equiv n_cU_{gg}$. Note that the non-condensate part is contributed by all momentum channels due to the quantum depletion, consistent with results from the dark state approach \cite{Jen}.

It is instructive to simplify Eq. (\ref{chi1}) further in the strong control field limit, i.e. $\Omega_2$ is much larger than atomic kinetic and interaction energies.\ It is straightforward to have following leading order results:
\begin{align}
\chi(\omega)
&=-\frac{d_0}{V}\sum_\k\int_{-\infty}^{\infty}d\tilde{\omega}i\tilde{G}^<(\k,\tilde{\omega})
\Big(\frac{\cos^2\phi}{\tilde{\omega}-\omega-\epsilon_{-}}\nonumber\\
&+\frac{\sin^2\phi}{\tilde{\omega}-\omega-\epsilon_{+}}\Big),
\label{chi2}
\end{align}
where $\cos\phi\equiv\sqrt{(\epsilon_+ +\Delta_1)/(\epsilon_+ - \epsilon_-)}$, and $\epsilon_\pm\equiv-\Delta_1+(\Delta_2\pm\sqrt{\Delta_2^2+4\Omega_2^2})/2$. Note that $\chi(\omega)$ obtained in this limit has no momentum ($\mathbf{q}$) dependence.

\section*{Single particle Green's function of a Luttinger liquid}

Here we demonstrate how to derive the single particle Green's function \cite{manyparticle} for a Luttinger liquid (LL).\ First, we use the density-phase representation of low-energy bosonic field operators, $\hat{\psi}_{g}^{\dagger}\left( x,t\right) =\sqrt{\hat{\rho}(x,t)}e^{-i\hat{\phi}(x,t)}$, for the Luttinger liquid \cite{LL}, where $\hat{\rho}(x,t)$ and $\hat{\phi}(x,t)$ are density and phase fluctuation operators. The single particle Green's function becomes
\begin{align}
&iG^<_{LL}(0,0;x,t) \nonumber\\
 & =\left\langle \hat{\psi}_{g}^{\dag}(x,t)\hat{\psi}_{g}(0,0)\right\rangle\theta(t)
,\nonumber\\
& \simeq\sqrt{\hat{\rho}(x)\hat{\rho}(0)}\left\langle e^{-i\hat{\phi}(x,t)}e^{i\hat{\phi}(0,0)}\right\rangle\theta(t) ,\nonumber\\
& \simeq n\left\langle e^{-i\hat{\phi}(x,t)}e^{i\hat{\phi}(0,0)}\right\rangle\theta(t) ,
\end{align}
where we have used the fact that the density fluctuation is suppressed and negligible for a repulsively interacting 1D gas \cite{LL}, and therefore only phase fluctuation is kept for the low energy effective theory. $\theta(t)$ is a Heaviside step function.\ The above equation can be simplified by using $e^{\hat{A}}e^{\hat{B}}=e^{\hat{A}+\hat{B}}e^{[  \hat{A},\hat{B}]  /2}$ if $\hat{A}$ and $\hat{B}$ are linear combination of bosonic operators and $[\hat{A},\hat{B}]$ is a complex number. We can also apply $\langle e^{\hat{A}}\rangle =e^{\langle \hat{A}^{2}\rangle /2}$ for the expectation on a bilinear Hamiltonian of $\hat{A}$. In the low energy limit, it has been shown that the effective Hamiltonian of a 1D bosonic gas can be described by a LL model, where the phase operator $\hat{\phi}(x,t)$ can be calculated within the periodic boundary condition \cite{LL},
\begin{align}
\hat{\phi}(x)&=\frac{1}{2}\sum_{q\neq 0}\bigg|\frac{2\pi}{qL\kappa}\bigg|^{1/2}\nonumber\\&\times e^{-a|q|/2}\text{sgn}(q)\left[e^{iqx}\hat{b}(q)+e^{-iqx}\hat{b}^\dagger(q)\right],
\label{phi_b}
\end{align}
where $\hat{b}(q)$ is the bosonic eigenstate field operator of the effective LL model. A positive length scale, $a$, is introduced as a cutoff length scale for the convergence of integrals, and $L$ is the system size.\ Here the LL parameter is denoted as $\kappa$, and the dispersion is $\omega(q)=|q|v$ with $v$ being the phonon velocity. Note that the density-phase representation in a LL is valid in the long wavelength limit $q\ll\rho^{-1}$. The exact values of $\kappa$ and $v$ should be determined by a more microscopic calculation or from the experimental measurement.

As a result, the single particle Green's function of the LL model can be calculated to be 
\begin{align}
&iG^<_{LL}(0,0;\r,t)\nonumber\\
&\simeq e^{\left[  \hat{\phi}(x,t),\hat{\phi}(0)\right]  /2}n\exp\left\{  -\frac{1}{2}\left\langle \hat{T}\left[  \hat{\phi}(x,t)-\hat{\phi}(0)\right]
^{2}\theta(t)\right\rangle \right\}  ,\nonumber\\
&\simeq e^{\left[  \hat{\phi}(x,t),\hat{\phi}(0)\right]  /2}n\exp\left\{  -\frac{1}{4\kappa}\ln\frac{x^2+(a+ivt)^2}{a^2}\right\}\theta(t),
\label{G_LL1}
\end{align}
where $\hat{T}$ is the time-ordered operator.\ This correlation function from LL model describes the behavior of quasi-long range order that has an exponent proportional to the interaction strength.\ The logarithmic function inside the exponential function is derived from (use $\omega_{q}=|q|v$ and Eq. (\ref{phi_b}))
\begin{align}
&\left\langle \hat{T}\left[  \hat{\phi}(x,t)-\hat{\phi}(0)\right]  ^{2}\right\rangle
\nonumber\\&=\frac{\theta(t)}{4}\sum_{q\neq0}\left\vert \frac{2\pi}{qL\kappa}\right\vert e^{-a\left\vert q\right\vert}\left[2-2e^{i(qx-|q|vt)}\right],\nonumber\\
& =\frac{\theta(t)}{2\kappa}\int_{0}^{\infty}dq\frac{e^{-aq}}{q}\left[ 2-e^{iq(x-vt)}-e^{-iq(x+vt)}\right],\nonumber\\
& =\frac{\theta(t)}{2\kappa}\ln\frac{x^2+(a+ivt)^2}{a^2},
\end{align}
which can be also found in Ref. \cite{Giamarchi}.\ The commutation relation in the prefactor of the Green's function in Eq. (\ref{G_LL1}) can be also calculated to be
\begin{align}
&\left[  \hat{\phi}(x,t),\hat{\phi}(0)\right]   &\nonumber\\
&=\frac{1}{4K}\int_{-\infty}^{\infty}\frac{e^{-a\left\vert q\right\vert }}{\left\vert q\right\vert }\left[
e^{i(qx-\left\vert q\right\vert vt)}-e^{-i(qx-\left\vert q\right\vert vt)}\right]  ,\nonumber\\
& =\frac{1}{4K}\log\frac{\left[  a+i\left(  x-vt\right)  \right]  \left[a-i\left(  x+vt\right)  \right]  }{\left[  a-i\left(  x-vt\right)  \right]
\left[  a+i\left(  x+vt\right)\right]}=0,
\end{align}
where in the last line we have taken the limit $a\ll |x|,v|t|$.

Finally, the full dynamical correlation function of a LL model (which is in general valid only in the low-energy and long wavelength limit of a 1D Bose gas) is then derived as
\begin{align}
&iG^<_{LL}(0,0;x,t)=\frac{n a^{1/(2\kappa)}}{[x^2+(a+ivt)^2]^{1/(4\kappa)}}.
\label{green}
\end{align}
\section*{The electric susceptibility of Luttinger liquid}

Here we proceed to calculate the electric susceptibility (i.e. EIT spectrum) of a LL. It is more instructive to consider the case of large control field limit (i.e. Eq. (\ref{chi2})) so that by inserting the result of Eq. (\ref{green}), we have
\begin{align}
&\chi_{LL}(\omega)\nonumber\\
&=\frac{id_{0}}{2\pi}\int_{-\infty}^{\infty}dk\int_{0}^{\infty}dt\int_{-\infty}^{\infty}dx\frac{n e^{ikx}a^{1/(2\kappa)}}{[x^2+(a+ivt)^2]^{1/(4\kappa)}}\nonumber\\
&\times\left[ \cos^{2}\phi e^{-i(\omega+\epsilon_-)t}+\sin^{2}\phi e^{-i(\omega+\epsilon_+)t}\right],\nonumber\\
&=id_{0}n\int_{0}^{\infty}dt\frac{a^{1/(2\kappa)}}{(a+ivt)^{1/(2\kappa)}}\left[ \cos^{2}\phi e^{-i(\omega+\epsilon_- )t}\right.
\nonumber\\&\left.+\sin^{2}\phi e^{-i(\omega+\epsilon_+ )t}\right],\nonumber\\
&=d_0n\Big(\frac{a}{v}\Big)^{1/(2\kappa)}\Gamma(1-\frac{1}{2\kappa})\times\nonumber\\&
\bigg[\frac{\cos^2\phi }{(\omega+\epsilon_{-})^{1-1/(2\kappa)}}+\frac{\sin^2\phi }{(\omega+\epsilon_{+})^{1-1/(2\kappa)}}\bigg],\label{LL_analytic}
\end{align}
where in the last line we have used the integral property \cite{integral},
\begin{align}
&\int_{0}^{\infty}dt\frac{e^{-i\omega t}}{(a+ivt)^b}=e^{a\omega/v}(iv)^{-b}(i\omega)^{-1+b}\Gamma(1-b,\frac{a\omega}{v}),
\end{align}
under the conditions that Re$[a]>0$, Im$[\omega]<0$, and Re$[v]\geq 0$.\ $\Gamma(s,x)$ is the incomplete gamma function which becomes the gamma function $\Gamma(s)$ when $a\rightarrow 0$.

Eq. (\ref{LL_analytic}) indicates a nontrivial power law dependence in the EIT spectrum, as shown in Fig. 1 from strong to weak interacting regimes ($\kappa=0.6-10$)\cite{explain}.\ The standard (noninteracting) EIT spectrum is similar to the weak interacting one (Fig. 1(d)) where the zero of the dispersion relation coincides with the transparency point ($\Delta_1=0$), and symmetric (anti-symmetric) absorption (dispersion) profile is retrieved around $\Delta_1=0$.\ When the interaction becomes stronger (smaller $\kappa$), the zero of dispersion relation moves away from the transparency point, and the EIT is destroyed for even stronger interaction.\ Interestingly, the EIT profile for $\kappa=1$ case (hard core boson limit) shows an inversion symmetry between absorption and dispersion profiles around $\Delta_1=0$, very different from the standard EIT spectrum in the noninteracting limit ($\kappa=10$).

\begin{figure}[t]
\centering\includegraphics[height=4.5cm, width=8.5cm]{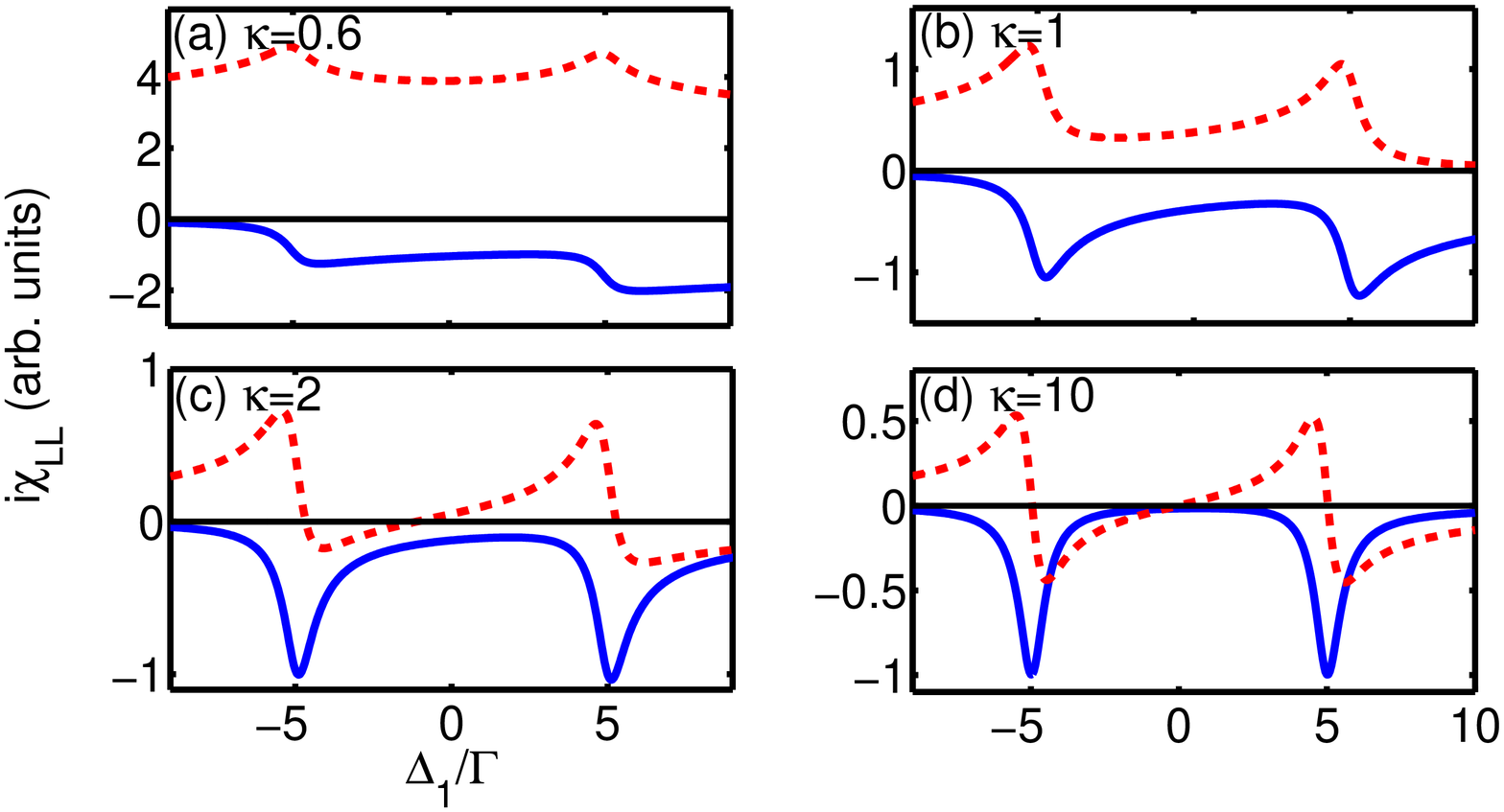}
\caption{(Color online) EIT profiles for a LL of $^{87}$Rb atoms in the large control field limit from Eq. (\ref{LL_analytic}). We take a static probe field ($\q,\omega=0$) and a resonant control field ($\Delta_2=0$) with Rabi frequency $\Omega_2=5~\Gamma$. The excited state is chosen as low-lying Rydberg transition of $|24\textrm{P}_{3/2}\rangle$ with $\Gamma^{-1}=28.3~\mu\text{s}$. The absorption (Re[$i\chi$], solid-blue) and dispersion (Im[$i\chi$], dash-red) profiles for (a) $\kappa=0.6$, (b) $1$, (c) $2$, and (d) $10$. The horizontal line guides the eye to the zero.}
\label{SM_1}
\end{figure}
\section*{Single particle Green's function for Mott state}

Here we show how to derive the single particle Green's function for a Mott state. At zero temperature, and in the deep Mott-insulating limit of average $n_0$ particle per site, we may assume only small number fluctuations about $n_0$ so that the Hilbert space can be truncated to three particle numbers per site only: $n_0-1$, $n_0$, and $n_0+1$. Within such three-state model \cite{Mott,three}, the original bosonic field operator at site ${\bf R}$ can be re-expressed to be: $\hat{g}^\dagger_\mathbf{R}(t)=\sqrt{n_0+1}\hat{t}^\dagger_{1,\mathbf{R}}(t)\hat{t}_{0,\mathbf{R}}(t)+\sqrt{n_0}\hat{t}^\dagger_{0,\mathbf{R}}(t)\hat{t}_{-1,\mathbf{R}}(t)$, where $\hat{t}_{\pm 1,\mathbf{R}}$ and $\hat{t}_{0,\mathbf{R}}$ are the raising and lowering operators. The conservation of total number of particles provides an additional constraint on the Hilbert space: $\sum_{\alpha=\pm1,0}\hat{t}_{\alpha,\mathbf{R}}^{\dag}(t)\hat{t}_{\alpha,\mathbf{R}}(t)=1$.

Since we are interested in the deep Mott insulator regime with little number fluctuation, we can apply the conservation of particle number shown above to have the following approximation \cite{Mott,three}: $\hat{t}_{0,\mathbf{R}}(t),\hat{t}_{0,\mathbf{R}}^{\dag}(t)\simeq 1-\frac{1}{2}\hat{t}_{1,\mathbf{R}}^{\dag}(t)\hat{t}_{1,\mathbf{R}}(t)-\frac{1}{2}\hat{t}_{-1,\mathbf{R}}^{\dag}(t)\hat{t}_{-1,\mathbf{R}}(t)$. As a result, the single particle correlation function can then be easily expressed to the quadratic order of small fluctuations to be 
\begin{widetext}
\begin{align}
\langle \hat{g}_\mathbf{R}^{\dag}(t)\hat{g}_{\mathbf{R}^{\prime}}(0)\rangle=
&(n_{0}+1)\langle \hat{t}_{1,\mathbf{R}}^{\dag}(t)\hat{t}_{0,\mathbf{R}}(t)\hat{t}_{0,\mathbf{R}^{\prime}}^{\dag}(0)\hat{t}_{1,\mathbf{R}^{\prime}}(0)\rangle
+n_{0}\langle\hat{t}_{0,\mathbf{R}}^{\dag}(t)\hat{t}_{-1,\mathbf{R}}(t)\hat{t}_{-1,\mathbf{R}^{\prime}}^{\dag}(0)\hat{t}_{0,\mathbf{R}^{\prime}}(0)\rangle 
\nonumber\\
&+\sqrt{n_{0}(n_{0}+1)}\left[\langle\hat{t}_{1,\mathbf{R}}^{\dag}(t)\hat{t}_{0,\mathbf{R}}(t)\hat{t}_{-1,\mathbf{R}^{\prime}}^{\dag}(0)\hat{t}_{0,\mathbf{R}^{\prime}}(0)\rangle
+\langle\hat{t}_{0,\mathbf{R}}^{\dag}(t)\hat{t}_{-1,\mathbf{R}}(t)\hat{t}_{0,\mathbf{R}^{\prime}}^{\dag}(0)\hat{t}_{1,\mathbf{R}^{\prime}}(0)\rangle\right],
\nonumber\\
\simeq &(n_{0}+1)\langle \hat{t}_{1,\mathbf{R}}^{\dag}(t)\hat{t}_{1,\mathbf{R}^{\prime}}(0)\rangle 
+n_{0}\langle \hat{t}_{-1,\mathbf{R}}(t)\hat{t}_{-1,\mathbf{R}^{\prime}}^{\dag}(0)\rangle +\sqrt{n_{0}(n_{0}+1)}
\nonumber\\
&\times\left[\langle \hat{t}_{1,\mathbf{R}}^{\dag}(t)\hat{t}_{-1,\mathbf{R}^{\prime}}^{\dag}(0)\rangle
+\langle \hat{t}_{-1,\mathbf{R}}(t)\hat{t}_{1,\mathbf{R}^{\prime}}(0)\rangle \right].
\label{G_Mott}
\end{align}
\end{widetext}
Within the MI state, it has been shown that we can diagonalize the effective Hamiltonian \cite{Mott} in the three state model, and change the lowering and raising operators by the quasi-hole and quasi-particle excitations: $\hat{\beta}_h$ and $\hat{\beta}_p$, i.e. $\hat{t}_{-1,\k}(t)=-B(\k)\hat{\beta}_{p,\k}^{\dag}(t)-A(\k)\hat{\beta}_{h,\k}(t)$, and $\hat
{t}_{1,-\k}^{\dag}(t)=A(\k)\hat{\beta}_{p,\k}^{\dag}(t)+B(\k)\hat{\beta}_{h,\k}(t)$, where 
\begin{align}
\begin{split}
A(\k)  & =\cosh(\frac{D_{\k}}{2}),~B(\k)=\sinh(\frac{D_{\k}}{2}),\\
\tanh(D_{\k})  & =\frac{-2\epsilon_{0}(\k)\sqrt{n_{0}(n_{0}+1)}}{U-\epsilon
_{0}(\k)(2n_{0}+1)},\\
\epsilon_{0}(\k)  & =2J\sum_{\alpha=1}^{3}\cos(k_{\alpha}a).
\end{split}
\end{align}
The corresponding eigenenergies of the particle and hole excitations are $\epsilon_{p,h}(\k)=\mp[\epsilon_{0}(\k)/2+\delta\mu]+\tilde{\omega}(\k)$, where $\tilde{\omega}(\k)$ $=$ $\sqrt{U^{2}-U\epsilon_{0}(\k)(4n_{0}+2)+\epsilon_{0}^{2}(\k)}/2$, and $\delta\mu= -3J$ for a 3D cubic lattice \cite{Mott}.

As a result, we can easily calculate a correlation function as following: 
\begin{align}
&\langle \hat{t}_{1,\mathbf{R}}^{\dag}(t)\hat{t}_{1,\mathbf{R}^{\prime}}(0)\rangle
\nonumber\\&=\frac{1}{L^{3}}\sum_{\k}B^{2}(\k)e^{-i\epsilon_{h}(\k)t}e^{-i\k\cdot(\mathbf{R}-\mathbf{R}^{\prime})}.
\end{align}
Similarly the single particle Green's function in Eq. (\ref{G_Mott}) becomes
\begin{align}
\langle\hat{g}_\mathbf{R}^{\dag}(t)\hat{g}_{\mathbf{R}^{\prime}}(0)\rangle=&\frac{1}{L^{3}}\sum_{\k}\left[\sqrt{(n_{0}+1)}B(\k)-\sqrt{n_{0}}A(\k)\right]^{2}
\nonumber\\&\times e^{-i\k\cdot(\mathbf{R}-\mathbf{R}^{\prime})}e^{-i\epsilon_{h}(\k)t}.
\end{align}

In the deep Mott regime, we have $U\gg J,$ and hence $A(\k)\to 1$, and $\langle \hat{g}_{R}^{\dag}(t)\hat{g}_{R^{\prime}}(0)\rangle=n_{0}e^{-i\epsilon_{h}(\k)t/\hbar}\delta_{R,R^{\prime}}$.\ The single particle Green's function for the Mott state could be further simplified to be (use $\tilde{\psi}_g(\k)$ as the Fourier transform of $\hat{\psi}_g(\r)=\sum_\mathbf{R}\hat{g}_\mathbf{R}\omega_\mathbf{R}(\r)$)
\begin{align}
i\tilde{G}^<(\k,t)&=\left\langle\tilde{\psi}^\dag_g(\k,t)\tilde{\psi}_g(\k,0)\right\rangle\theta(t),\nonumber\\
&=\sum_{\mathbf{R},\mathbf{R}'}\tilde{\omega}^*_\mathbf{R}(\k)\tilde{\omega}_{\mathbf{R}'}(\k)\left\langle\hat{g}^\dag_\mathbf{R}(t)\hat{g}_{\mathbf{R}'}(0)
\right\rangle\theta(t)\nonumber\\
&=\sum_\mathbf{R}\left\vert\tilde{\omega}_\mathbf{R}(\k)\right\vert^2 n_0e^{-i\epsilon_h(\k)t}\theta(t).
\end{align} 
Note that only hole excitation energy appears because the Green's function we need for EIT spectrum is time-ordered, i.e. an atom is excited from the ground state ($|g\rangle$) to the excited state ($|e\rangle$) by the probe field, leaving a hole excitation inside the strongly interacting system.

\section*{Single particle Green's function for superfluid case of two-component Fermi gases}

Here we show how to derive the single particle Green's function for a BCS superfluid state of two-component Fermi gases.\ The single particle Green's function in this context is defined as
\begin{align}
-i\tilde{G}_{BCS}^<(\k,t)&=\left\langle\hat{g}^\dag_{\k,\uparrow}(t)\hat{g}_{\k,\uparrow}(0)\right\rangle\theta(t),
\end{align} 
where we define the original ground state (in the EIT $\Lambda$ scheme) to be spin up. The other state of spin down is assumed not directly involved in the EIT experiment. To evaluate the above expectation, we can express the ground state in terms of Bogoliubov quasi-particles (denoted by $\hat{\alpha}$ and $\hat{\beta}$) of Cooper pairs in the superconducting state,
\begin{align}
\hat{g}_{\k,\uparrow}=\cos\theta_\k\hat{\alpha}_\k+\sin\theta_\k\hat{\beta}_{-\k}^\dagger,
\end{align}
where $\sin^2\theta_\k=(1-\xi_\k/E_\k)/2$.\ Here $E_\k=\sqrt{\Delta_{S}^2+\xi_\k^2}$ is the excitation energy of the quasi-particles with $\xi_k\equiv\k^2/(2m)-\mu$. $\mu$ is the chemical potential and is determined by the particle density. The Green's function can then be obtained to be
\begin{align}
-i\tilde{G}^<_{BCS}(\k,t)&=\left\langle\left(\cos\theta_\k \hat{\alpha}_{\k}^\dag e^{iE_{\k}t}+\sin\theta_\k \hat{\beta}_{-\k}e^{-iE_{\k}t}\right)\right.\times\nonumber\\
&\left.\left(\cos\theta_\k \hat{\alpha}_{\k}+\sin\theta_\k \hat{\beta}_{-\k}^\dag\right)\right\rangle_{H}\theta(t),\nonumber\\
&=\sin^2\theta_\k e^{-iE_\k t}\theta(t).
\end{align}


\begin{thebibliography}{99}
\bibitem{B} In this paper, we just consider low-lying Rydberg states ($n'$P levels, $n'<30$) \cite{Rydberg,30P} of alkali atoms.\ The van der Waals coefficient $|\text{C}_6|$ for $30$P-$30$P rubidium asymptotes is in the range of $(0.26-4.7)\times 10^{16}$ a.u. \cite{C6}, giving us the blockade radius $\text{R}_\text{c}=0.88-1.29$ $\mu$m \cite{superatom,superatom2} for a density $n=10^{14}$ $\text{cm}^{-3}$ with the effective excitation Rabi frequency $\Omega_1=2\pi\times100$ kHz.\ This radius provides an upper estimation of the dipole blockade strength, which can be further reduced by decreasing $\Omega_1$. Note that the probe field has a fundamental limit of single photon shot noise which can be overcome by more experimental trials. As a result, we neglect the blockade in our theory.
\end{thebibliography}
\begin{thebibliography}{99}
\bibitem{explain} Note that $\kappa$ can be smaller than one when considering a long-ranged repulsive interaction.
\end{thebibliography}
\end{document}